\documentclass[reprint,prab]{revtex4-2}

\usepackage[colorlinks=true, allcolors=blue]{hyperref}
\usepackage{graphicx}
\usepackage{dcolumn}
\usepackage{bm}
\usepackage{siunitx}
\usepackage{xcolor}
\usepackage{braket}
\usepackage{mathtools}
\usepackage{amsmath}
\usepackage{leftidx}
\usepackage{wasysym}
\usepackage{float}
\usepackage{cleveref}
\Crefname{equation}{Eq.}{Eqs.}

\usepackage{physics}

\usepackage{mypackage}
\usepackage{numericalvalues}
\usepackage{beamdivergence}

\usepackage[T1]{fontenc}
\DeclareSIUnit[number-unit-product = {}]{\inch}{\text{\textquotedbl}}

\usepackage{amsfonts} 
\DeclareMathSymbol{\shortminus}{\mathbin}{AMSa}{"39}

\graphicspath{{images/}}

\begin{document}

\title{Measurements of undulator radiation power noise and comparison with \textit{ab initio} calculations}

\author{Ihar Lobach}
\email{ilobach@uchicago.edu}
\affiliation{The University of Chicago, Department of Physics, Chicago, IL
60637, USA}
\author{Sergei Nagaitsev}
\altaffiliation[Also at ]{The Enrico Fermi Institute, The University of Chicago,
Chicago, Illinois 60637, USA}
\author{Valeri Lebedev}
\author{Aleksandr Romanov}
\author{Giulio Stancari} 
\author{Alexander Valishev} 
\affiliation{Fermi National Accelerator Laboratory, Batavia, IL 60510, USA}%
\author{Aliaksei Halavanau}
\author{Zhirong Huang}
\affiliation{SLAC National Accelerator Laboratory, Stanford University, Menlo
Park CA 94025, USA}
\author{Kwang-Je Kim}
\altaffiliation[Also at ]{The Enrico Fermi Institute, The University of Chicago,
Chicago, Illinois 60637, USA} \affiliation{Argonne National Accelerator
Laboratory, Lemont, IL 60439, USA}%

\begin{abstract}
Generally, turn-to-turn fluctuations of synchrotron radiation power in a
storage ring depend on the 6D phase-space distribution of the electron bunch. This
effect is related to the interference of fields radiated by different electrons.
Changes in the relative electron positions and velocities inside the bunch
result in fluctuations in the total emitted energy per pass in a synchrotron radiation source. This effect has been previously described assuming constant and equal electron velocities before
entering the synchrotron radiation source. In this paper, we present a
generalized formula for the fluctuations with a non-negligible beam divergence.
Further, we corroborate this formula in a dedicated experiment with undulator
radiation in the Integrable Optics Test Accelerator (IOTA) storage ring at Fermilab. Lastly,
possible applications in beam instrumentation are discussed. 
\end{abstract}

\maketitle


\section{Introduction}

Full understanding of the radiation generated by accelerating charged particles
is crucial for accelerator physics and electrodynamics in general. The
predictions of classical electrodynamics for pulse-by-pulse average
characteristics of synchrotron radiation, such as the total radiated power, spectral
composition, angular intensity distribution and brightness
\cite{kim2017synchrotron}, are supported by countless observations. In fact,
they are confirmed every day by routine operations of synchrotron radiation
user facilities. On the other hand, the pulse-to-pulse statistical fluctuations
of synchrotron radiation have not been studied at the same level of detail yet,
although substantial progress has been made in the past few decades. The turn-to-turn intensity fluctuations of incoherent spontaneous bending-magnet,
wiggler, and undulator radiation in storage rings have been studied in
Refs.~\cite{teich1990statistical,sajaev2000determination,sajaev2004measurement,lobach2020PRAB,sannibale2009absolute,catravas1999measurement},
both theoretically and experimentally. The statistical properties of the
Free-Electron Laser (FEL) radiation have been studied in
Refs.~\cite{kim1997start,benson1985shot,saldin2013physics,pellegrini2016physics,becker1982photon,becker1983fully,becker1983photon}.
Moreover, Refs.~\cite{chen2001observation,chen1999photon} claimed to observe a non-classical sub-Poissonian photon statistics in the seventh coherent
spontaneous harmonic of an FEL, although it could have been an instrumentation effect
\cite{park2019investigation}. In any case, more experimental studies into the
statistical properties of synchrotron radiation are needed. In this paper, we describe our observation of
turn-to-turn power fluctuations of incoherent spontaneous undulator radiation in
the Integrable Optics Test Accelerator (IOTA) storage ring at Fermilab \cite{antipov2017iota}. Also, we extend the existing theoretical description
\cite{sannibale2009absolute,lobach2020PRAB} of such fluctuations. Namely, in
Refs.~\cite{sannibale2009absolute,lobach2020PRAB}, only the effect of spatial
distribution of the electrons inside the bunch on the turn-to-turn fluctuations
is considered. However, in general, the distribution of electron velocities
affects the fluctuations as well. We present a generalized formula for the
fluctuations in the case of non-negligible beam divergence in this paper.

Fluctuations and noise do not always degrade the results of an experiment. There
are numerous examples when noise is used to measure the parameters of a specific
system, or even fundamental constants. Some examples are the pioneering
determination of the elementary charge $e$ by the shot noise
\cite{hull1925determination}, and the determination of the Boltzmann constant
$k_B$ by the Johnson-Nyquist noise \cite{johnson1928thermal}. Another example,
relevant to the field of accelerator physics, is the use of Schottky noise
pick-ups in storage rings
\cite{boussard1986schottky,van1989diagnostics,caspers20074}
to measure transverse rms emittances, momentum spread, number of particles, etc.
In fact, we will see that the fluctuations in synchrotron radiation are similar
to Schottky noise. Both effects are related to the existence of discrete
point-like charges as opposed to a continuos charge fluid. Therefore,
measurements of electron bunch parameters via synchrotron radiation fluctuations
have been reported too. They were mostly focused on the longitudinal bunch length \cite{sannibale2009absolute,sajaev2004measurement,
sajaev2000determination}. Reference~\cite{catravas1999measurement} reported an
order-of-magnitude measurement of a transverse emittance. In this paper, we
present one example of a measurement of an unknown small vertical emittance of a
flat beam in IOTA, given a known horizontal emittance, a longitudinal bunch
shape, and ring focusing functions, using our new formula for the fluctuations. Taking beam divergence into account is critical in this specific measurement. For
more results regarding beam diagnostics via fluctuations in IOTA, we refer
the reader to our separate publication \cite{lobach2020furjointprl}.

\section{\label{sec:theory}Theoretical description} Let us assume that we have a detector that can measure the number of detected synchrotron radiation photons
$\N$ at each revolution in a storage ring. Then, according to
\cite{lobach2020PRAB,kim2017synchrotron,teich1990statistical}, the variance of this number can be expressed as 
\begin{equation}\label{eq:varN_from_book}
    \var{\N}=\av{\big(\N-\av{\N}\bigr)^2}=\av{\N}+\frac{1}{M}\av{\N}^2,
\end{equation}
\noindent where the linear term represents the photon shot noise, related to the
quantum discrete nature of light. This effect would exist even if there was only
one electron. Indeed, the electron would radiate photons with Poisson
statistics~\cite{glauber1963quantum,glauber1963coherent,glauber1951some}. The
quadratic term corresponds to the interference of fields radiated by different
electrons. Changes in relative electron positions and velocities, inside the
bunch, result in fluctuations of the radiation power, and, consequently, of the
number of detected photons. In a storage ring, the effect arises because of betatron motion, synchrotron motion, radiation induced diffusion, etc. The
dependence of $\var{\N}$ on the electron bunch parameters is introduced through
the parameter $M$, which will be called the ``number of coherent modes'', following
the nomenclature of \cite{kim2017synchrotron,teich1990statistical,lobach2020PRAB}.
In \cite{lobach2020PRAB}, we derived an equation for $M$ for an electron bunch
with a Gaussian transverse density distribution and an arbitrary longitudinal density
distribution, assuming an rms bunch length much longer than the radiation
wavelength and a negligible electron beam divergence.  In this paper, we present
an equation for $M$ extended to an arbitrary beam divergence,
\newcommand{\bigP}{\mathcal{P}_k(\xpv,\phiv_1-\phiv_2)}
\begin{widetext}
\begin{equation}\label{eq:Mgeneral}
    \frac{1}{M} = (1-1/\Ne)
    \frac{\sqrt{\pi}}{\szeff}
    \frac{
        \int 
        \dd{k} \dd[2]{\phiv_1} \dd[2]{\phiv_2} \dd[2]{\xpv}
        \bigP
        \mathcal{I}_k(\phiv_1,\xpv)
        \mathcal{I}_k^{*}(\phiv_2,\xpv)
        }{
            \None^2
            }
    ,
\end{equation}
\noindent with
\begin{equation}
    \bigP=
    \frac{1}{4\pi\sxp\syp}
    e^{
        -\frac{(\xp)^2}{4\sxp^2}
        -\frac{(\yp)^2}{4\syp^2}
    }
    e^{
        -ik\dx(\f{1x}-\f{2x})\xp
        -ik\dy(\f{1y}-\f{2y})\yp   
    }
    e^{
        -k^2\Sx^2(\f{1x}-\f{2x})^2
        -k^2\Sy^2(\f{1y}-\f{2y})^2
    },
\end{equation}
\end{widetext}
\begin{equation}
    \mathcal{I}_k(\phiv,\xpv) = 
    \sum\limits_{s=1,2}
    \etaf{\phiv}
    \Ef{\phiv}{}{s}
    \Ef{\phiv-\xpv}{*}{s},
\end{equation}
\begin{equation}\label{eq:NoneDef}
    \None = \sum\limits_{s=1,2} \int \dd{k}\dd[2]{\phiv}\etaf{\phiv}\bigl|\Ef{\phiv}{}{s}\bigr|^2,
\end{equation}
\noindent where $s=1,2$ indicates the polarization component, $\Ne$ is the
number of electrons in the bunch, $k = 2\pi/\lambda$ is the magnitude of the wave
vector; $\phiv = (\f{x},\f{y})$, $\phiv_1=(\f{1x},\f{1y})$ and $\phiv_2=(\f{2x},\f{2y})$ represent angles of
direction of the radiation in the paraxial approximation. Hereinafter, $x$ and
$y$ refer to the horizontal and the vertical axes, respectively, and
\begin{equation}\label{eq:kappaDef}
\szeff = 1/\bigl(2\sqrt{\pi}\int\rho^2(z)\dd{z}\bigr),
\end{equation}
\noindent where $\rho(z)$ is the electron bunch longitudinal density
distribution function, $\int\rho(z)\dd{z}=1$, and $\szeff$ is equal
to the rms bunch length $\sz$ for a Gaussian bunch; $\xpv=(\xp,\yp)$ represents the
direction of motion of an electron at the radiator center, relative to a reference electron; $\sxp$ and
$\syp$ are the rms beam divergences, $\sxp^2 = \gx \ex + \Dxp^2\sp^2$, $\syp^2 =
\gy\ey$; $\Sx^2 = \ex/\gx+(\gx\Dx+\Dxp\ax)^2\bx\ex\sp^2/\sxp^2$, $\Sy^2 =
\ey/\gy$, $\dx = (\ax\ex-\Dx\Dxp\sp^2)/\sxp^2$, $\dy = \ay/\ey$, where $\ax$,
$\bx$, $\gx$, $\ay$, $\by$, $\gy$ are the Twiss parameters of an uncoupled focusing optics in the synchrotron
radiation source; $\Dx$, $\Dxp$ are the horizontal dispersion and its derivative, and the
vertical dispersion is assumed to be zero; $\ex$, $\ey$ are the unnormalized rms
emittances; $\sp$ is the relative rms momentum spread. The following two useful relations
exist, $\sx^2 = \Sx^2 + \sxp^2\dx^2$, $\sy^2 = \Sy^2 + \syp^2\dy^2$, where $\sx$ and $\sy$ are the transverse rms beam sizes. The complex
radiation field amplitude $\Ef{\phiv}{}{s}$, generated by a reference
electron, is given by the following expression, see
\cite{glauber1963coherent,lobach2020PRAB},\cite[p.~38]{kim2017synchrotron},
\begin{equation}\label{eq:Adef}
    \Ef{\phiv}{}{s} = \sqrt{\frac{\alpha k}{2(2\pi)^3}}\int \dd{t} \epolar\cdot\vRef{t}
    e^{ick t - i\kv\cdot\Rref{t}}.
\end{equation}
\noindent where $\kv =
k(\f{x},\f{y},1-\f{x}^2/2-\f{y}^2/2)$, $\alpha$ is the fine-structure constant, $\epolar$ is the considered polarization vector ($s=1,2$), $\Rref{t}$ is the trajectory of the
reference electron in the synchrotron radiation source, $\vRef{t}$ is the
velocity of the reference electron as a function of time, $c$ is the speed of light. The electrons are assumed to be ultrarelativistic, $\gamma\gg1$, where $\gamma$ is the Lorentz factor.

The parameter $\None$ in \Cref{eq:Mgeneral} is the average number of detected photons per turn for a single
electron (s.e.) circulating in the ring. We consider the case of an incoherent radiation ($\sz k\gg1$). Therefore, the
average number of detected photons for the entire bunch can be obtained as
\begin{equation}\label{eq:avNeq}
    \av{\N} = \Ne \None.
\end{equation}

The integrals in \Cref{eq:Mgeneral,eq:NoneDef} are taken from minus to plus infinity over all integration variables except for $k$, which goes from zero to plus infinity. The spectral sensitivity and the aperture of the detector are assumed to be included in the detection efficiency $\etaf{\phiv}$, which is a
function of polarization, $k$, and $\phiv$ for that reason.

The derivation of
\Cref{eq:Mgeneral} is largely analogous to \cite{lobach2020PRAB} and is outlined
in Appendix~\ref{app:beamdiv}. Appendix~\ref{app:gaussianlightbeam} provides an illustrative closed-form expression for $M$, based on \Cref{eq:Mgeneral} in the
approximation of a Gaussian spectral-angular distribution of the radiation.

In IOTA, we study undulator radiation, because the quadratic term in
\Cref{eq:varN_from_book}, sensitive to bunch parameters, is larger for
undulators and wigglers than it is for dipole magnets \cite{lobach2020PRAB}. The complex field amplitude
$\Ef{\phiv}{}{s}$, generated by a single electron, can be numerically calculated by our computer code \cite{wigradrepo}, based on the equations from
\cite{clarke2004science}, or by using the SRW package
\cite{chubar2013wavefront}. Then, the integrals in \Cref{eq:Mgeneral,eq:NoneDef} can be
calculated by a Monte-Carlo algorithm. Our \CC~code with Python bindings for
calculation of \Cref{eq:Mgeneral,eq:NoneDef} is provided in the repository
\cite{furrepo}.

\section{\label{sec:apparatus}Apparatus}

\begin{figure}[!h]
    \includegraphics[width=0.7\columnwidth]{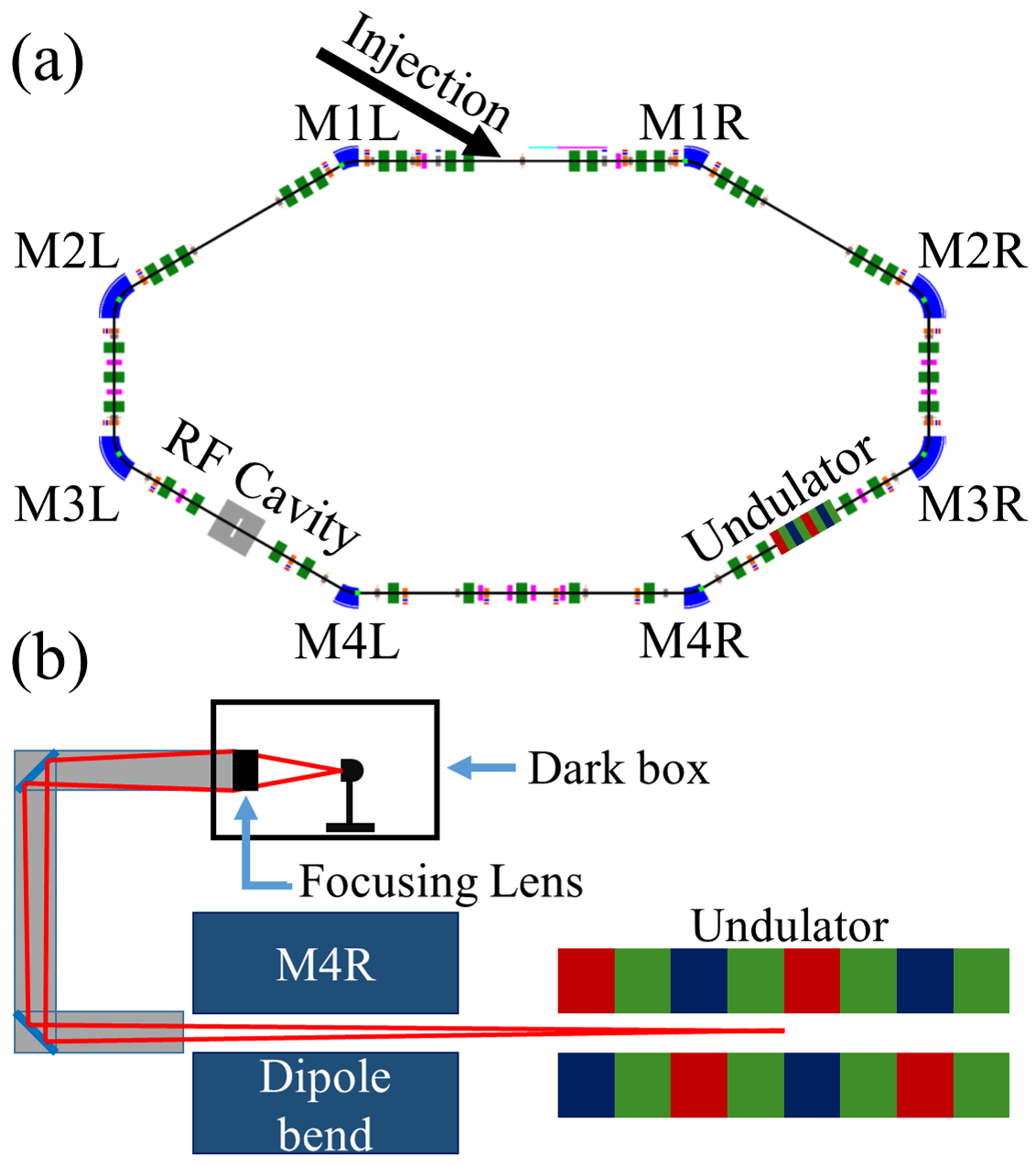}
    \caption{\label{fig:iota_detector_layout} (a) Layout of IOTA. (b) Light path
    from the undulator to the detector  (not to scale).}
\end{figure}

In our experiment, a single electron bunch circulated in the IOTA ring, see
Fig.~\ref{fig:iota_detector_layout}(a),
with a revolution period of $\valrevolutionPeriod$ and the beam energy of $\valEbeamDimLess \pm \valEbeamerror$. We studied two transverse focusing
configurations in IOTA: (1) strongly coupled, resulting in approximately equal transverse mode emittances
and (2) uncoupled, resulting in two drastically different emittances.
Henceforth, we will refer to the beams in these configurations as ``round'' and ``flat'' beams, respectively. In both
cases, the bunch length and the emittances depend on the beam current due to
intra-beam scattering \cite{bjorken1982intrabeam, IBS_Nagaitsev}, beam interaction with its environment \cite{haissinski1973exact}, etc.
The longitudinal bunch density distribution $\rho(z)$ was measured and recorded by
a high-bandwidth wall-current monitor \cite{WGM}. It was not exactly Gaussian,
but this fact was properly accounted for by \Cref{eq:kappaDef} for $\szeff$, which
works for any longitudinal bunch shape. The IOTA rf cavity operated at
\SI{30}{MHz} (4th harmonic of the revolution frequency) with a voltage amplitude of about $\valVrf$.
The rms momentum
spread $\sp$ was calculated from the known rf voltage amplitude, the design ring parameters and the measured rms bunch length $\sz$. In our experiments, the relation was
\begin{equation}
    \sp \approx \valszTospDimLess\times\sz[\mathrm{cm}].
\end{equation}
\noindent It is an approximate equation, because of the bunch-induced rf voltage (beam loading) and a
small deviation of $\rho(z)$ from the Gaussian shape. However, the effect of
$\sp$ in \Cref{eq:Mgeneral} in IOTA was almost negligible. Therefore, such
estimation was acceptable.

For the round beam, the IOTA transverse focusing
functions (4D Twiss functions) were chosen to produce approximately equal mode emittances at zero
beam current, $\e_1 \approx \e_1 \approx \SI{12}{nm}$ (rms, unnormalized).
It was empirically confirmed that they remained equal at all beam currents
with a few percent precision.
The expected zero-current emittances for a flat beam were $\ex
\approx \SI{50}{nm}$, $\ey \gtrsim \SI{0.33}{p m}$ (set by the quantum excitation in a perfectly uncoupled ring). The expected zero-current rms bunch length and the rms momentum spread for both round and flat beams were $\sz=\valszZeroCur$,
$\sp=\valspZeroCur$.  In our experiment, the electron beam sizes were monitored
and recorded by visible synchrotron light image monitors (SLMs) \cite{kuklev2019synchrotron} in
seven dipole bend locations, at M1L-M4L and at M1R-M3R, see
Fig.~\ref{fig:iota_detector_layout}(a).
The smallest reliably resolvable emittance by the SLMs in our experiment configuration was about $\SI{20}{nm}$.
\begin{figure}[!h]
    \includegraphics[width=1.0\columnwidth]{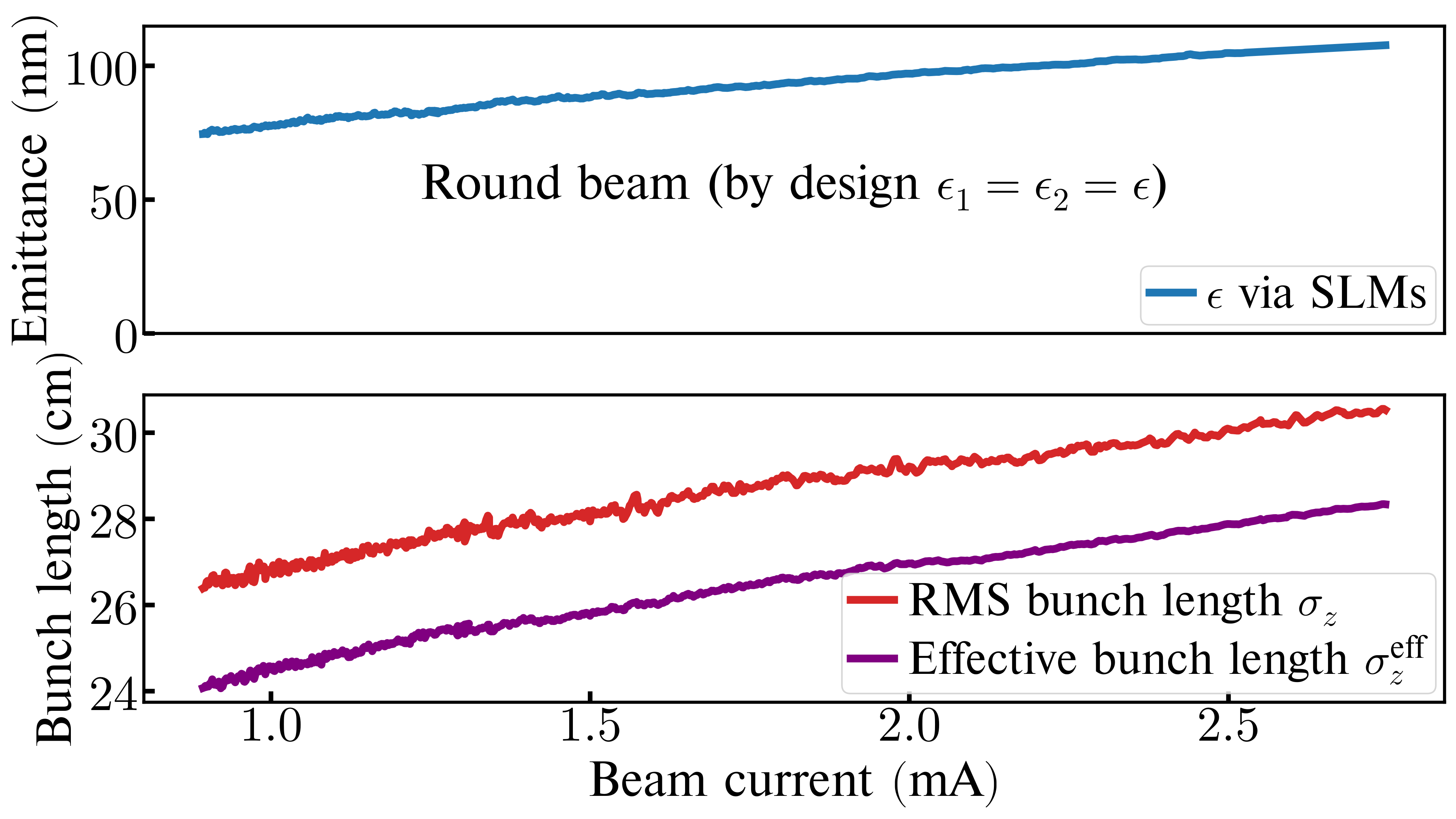}
    \caption{\label{fig:ex_ey_round_beam}
    Mode emittances ($\e_1=\e_2=\e$) and bunch lengths (rms and
    effective) of a round beam in IOTA as functions of beam current. SLMs had a
    monitor-to-monitor variation of $\pm\valSLMroundErr$ (not shown) in the measurement of $\e$. }
\end{figure}
Figure~\ref{fig:ex_ey_round_beam} illustrates the bunch parameters of the round
beam as a function of current. Below, we will present measurements with a flat
beam at only one value of beam current, $\valIbeamND$, measured with a direct-current current transformer (DCCT). The parameters of the flat
beam at this current value were $\sz=\valszND$, $\szeff=\valszeffND$,
$\sp=\valdppND$. The horizontal emittance was $\ex=\valexND$, as measured by the
SLMs with a monitor-to-monitor variation of $\pm\valSLMflatErr$. The small vertical
emittance of the flat beam was unresolvable by the SLMs. However, in
Sec.~\ref{sec:discussion}, we will demonstrate how it can be reconstructed using the fluctuations measurements.

At the center of the undulator, in the uncoupled optics, the Twiss parameters were $\bx=\valbetax$,
$\by=\valbetay$, $\ax=\valalphax$, $\ay = \valalphay$, the horizontal dispersion
$\Dx=\valDx$, its derivative $\Dxp=\valdDxND$.
The strongly coupled optics was created
from the uncoupled optics by changing the current in one skew-quad located at a zero dispersion location.
The coupling parameter $u$ \cite{lebedev2010betatron} was about \SI{0.5}{} everywhere in the ring. Therefore, the following is correct for the coupled case 4D Twiss functions, $\beta_{1x}\approx\beta_{2x}$, $\beta_{1y}\approx\beta_{2y}$. Moreover, their sums, $\beta_{1x}+\beta_{2x}$, $\beta_{1y}+\beta_{2y}$, were approximately equal to the Twiss beta functions in the uncoupled case, $\beta_x$, $\beta_y$. Equation~\eqref{eq:Mgeneral} assumes an uncoupled optics. However, this specific strongly coupled optics used in IOTA can be approximated by the uncoupled optics with equal horizontal and vertical emittances $\ex=\ey=\e$. More specifically, what is used in the derivation of \Cref{eq:Mgeneral} (see Appendix~\ref{app:beamdiv}) is the 6D phase-space distribution of the electrons, \Cref{eq:psixdim}. This distribution, for the round beam, when calculated using the approximation of uncoupled optics with equal emittances $\ex=\ey=\e$, and the distribution, calculated using the exact 4D Twiss functions and equal mode emittances, $\e_1=\e_2=\e$, are almost indistinguishable. This property was intentionally included in the initial design of the coupled optics in IOTA.

The undulator strength parameter
(peak) was $\Ku=\valKu$ with the number of periods $\Nu=\valNundPer$ and the period length $\lamu=\SI{5.5}{cm}$, the total length of the undulator was $\Lu=\Nu\lamu=\SI{58}{cm}$. A photodetector was installed in a dark box above the
M4R dipole magnet, see Fig.~\ref{fig:iota_detector_layout}(b). The light produced in the undulator was directed to the dark box by a
system of two mirrors ($\diameter \SI{2}{\inch}$). Then, it was focused by a lens
($\diameter \SI{2}{\inch}$, focal distance $F=\SI{150}{mm}$) into a spot, smaller
than the sensitive area of the detector ($\diameter \SI{1.0}{mm}$). The lens was $\valzobs$ away from the center of the undulator. Because of
the two round mirrors, which are at $\SI{45}{\degree}$ to the direction of
propagation of the radiation, the angular aperture takes an elliptical shape
with the vertical axis smaller than the horizontal by a factor of $\sqrt{2}$.
Namely, the horizontal and the vertical semi-axes were $\valsemiApertureX$ and $\valsemiApertureY$, respectively. The measurements were performed in the vicinity of the fundamental of the undulator
radiation, $\lambda_1=\lamu(1+\Ku^2/2)/(2\gamma^2)=\vallambdaone$.
As a photodetector we used an InGaAs PIN
photodiode \cite{hamamatsu_photodiode}, which has a high quantum efficiency ($\approx\SI{80}{\percent}$) around the fundamental.

\begin{figure}[!h]
    \includegraphics[width=1.0\columnwidth]{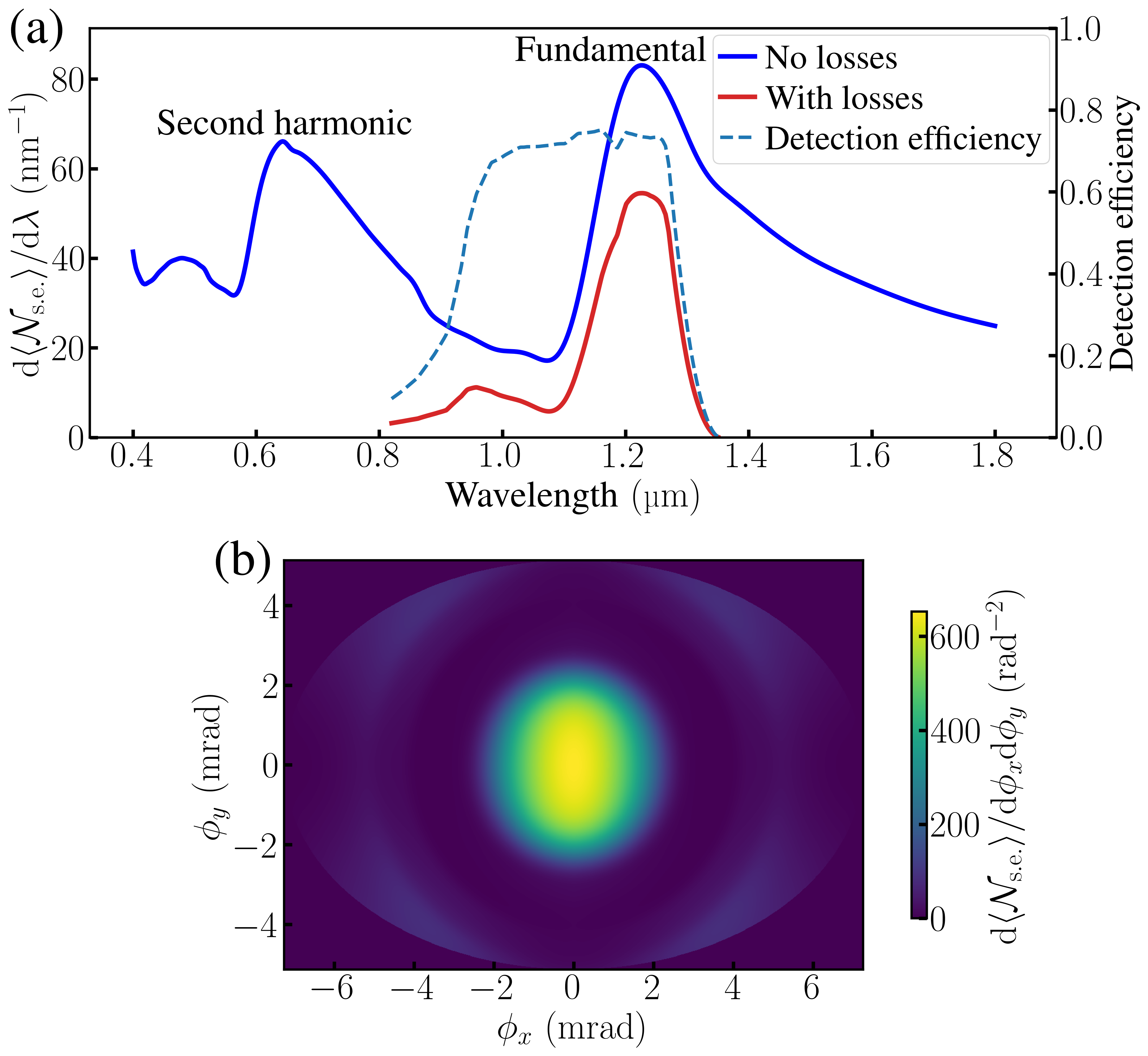}
    \caption{\label{fig:spectral_angular_distribution}
    (a) Spectral distribution of the average number of detected photons per turn for a single electron (s.e.)
    assuming no losses (blue) and accounting for the detection efficiency of the
    system (red). Also, the detection efficiency (dashed, right vertical scale). (b) Angular distribution of the number of detected photons
    accounting for the detection efficiency of our system. Both (a) and (b) are calculated for
    an elliptical aperture with the horizontal and the vertical semi-axes
    $\valsemiApertureX$ and $\valsemiApertureY$, respectively.}
\end{figure}

Using the elliptical angular aperture mentioned above and the manufacturers'
specifications for the spectral transmission of the vacuum chamber window at the
M4R dipole magnet, the two mirrors, the focusing lens, and the quantum
efficiency of the InGaAs photodiode, we constructed the detection efficiency
function $\etaf{\phiv}$ for our system. The lens's spectral transmission had to
be linearly extrapolated for a small interval outside of the range provided in the
manufacturer's specifications. There were no free adjustable parameters. We
calculated the field amplitude $\Ef{\phiv}{}{s}$, generated by a single
electron, for the parameters of our undulator on a 3D grid ($k$, $\phi_x$,
$\phi_y$) with our code \cite{wigradrepo}.
Figure~\ref{fig:spectral_angular_distribution}(a) shows the simulated spectrum,
where the intensity is integrated over the elliptical aperture,
\begin{equation}
    \dv{\None}{k}=\sum\limits_{s=1,2}\int \dd[2]{\phiv}\etaf{\phiv}\abs{\Ef{\phiv}{}{s}}^2.
\end{equation}

\begin{figure*}[t]
    \includegraphics[width=\textwidth]{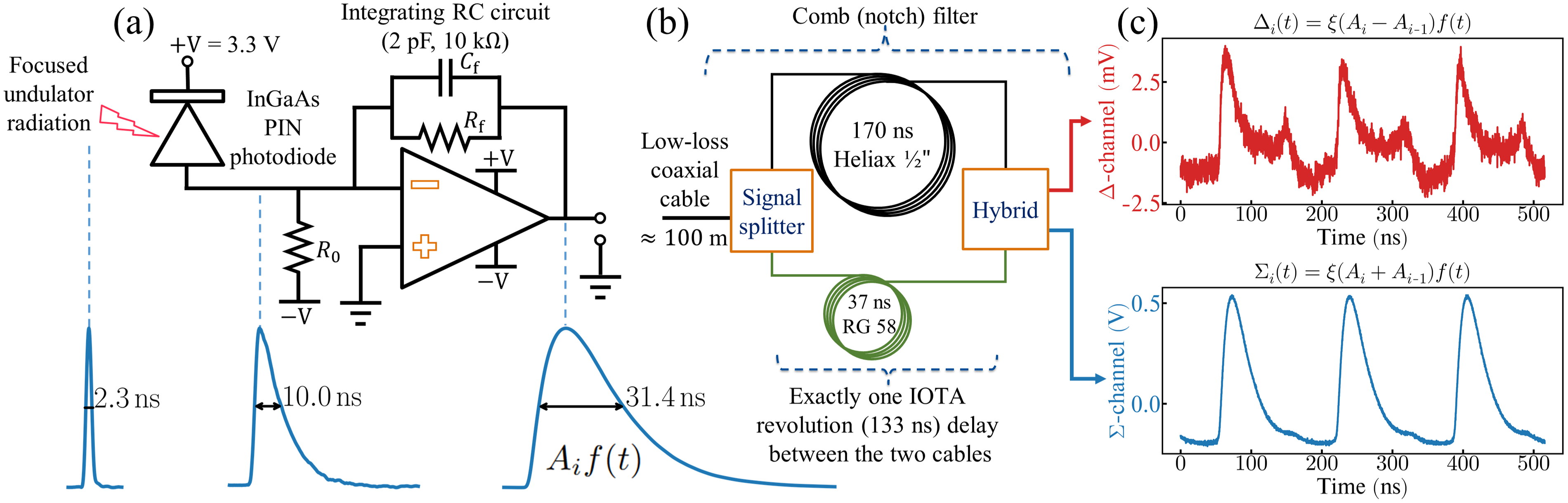}
    \caption{\label{fig:integrator_and_comb_filter} (a) Photocurrent integrator
    circuit. (b) Comb filter. (c) Sample waveforms of hybrid outputs (\D- and
    \S-channels).}
\end{figure*}

The blue line is calculated for an ideal detection system, where
$\etaf{\phiv}=1$ inside the elliptical aperture, and zero outside. The red line
is calculated with $\etaf{\phiv}$, constructed using the manufacturers'
specifications of the optical elements in our system. This $\etaf{\phiv}$ is equal to zero outside of the elliptical aperture. Whereas, inside, it is equal to the detection efficiency of our system. In our case, the detection efficiencies for the horizontal and vertical polarizations ($s=1,2$) are practically the same. Only the reflectance of the mirrors is slightly polarization dependent (under \SI{1}{\percent} difference). Moreover,  the radiation is dominated by the horizontal polarization (about \SI{96.5}{\percent}). The dashed line in Fig.~\ref{fig:spectral_angular_distribution}(a) is the detection efficiency of our system for the horizontal polarization.     
Figure~\ref{fig:spectral_angular_distribution}(b) shows the angular distribution
with $\etaf{\phiv}$ of our system,
\begin{equation}
    \frac{\dd[2]{\None}}{\dd\phi_x\dd\phi_y} = \sum\limits_{s=1,2}\int \dd{k}\etaf{\phiv}\abs{\Ef{\phiv}{}{s}}^2.
\end{equation}

With the spectral properties of all optical elements in the system taken into account, the spectral width of the radiation was $\valfwhm$ (FWHM), and the  angular size was $\approx\SI{2}{mrad}$, which could be fully transmitted through the $\diameter \SI{2}{\inch}$ optical system.

Figure~\ref{fig:integrator_and_comb_filter} illustrates our full photodetector
circuit. First, the radiation pulse is converted into a photocurrent pulse by
the photodiode, see Fig.~\ref{fig:integrator_and_comb_filter}(a). Then, the
photocurrent pulse is integrated by an op-amp-based RC integrator, which outputs
a longer pulse with a voltage amplitude that can be easily measured. The op-amp \cite{opamp} was capable of driving the 50-$\Omega$ input load
of a fast digitizing scope, located $\approx\SI{100}{m}$ away. The resistor
$R_0=\SI{580}{\kilo\ohm}$ in the circuit in
Fig.~\ref{fig:integrator_and_comb_filter}(a) was used to remove the offset in
the integrator output signal (about $\SI{0.3}{V}$), produced by the op-amp input
bias current and the photodiode leakage current. The output voltage pulse of the
integrator at the $i$th IOTA turn can be represented as $A_i f(t)$, where $A_i$ is
the signal amplitude at the $i$th turn and $f(t)$ is the average signal for one
turn, normalized so that its maximum value is 1, see
Fig.~\ref{fig:integrator_and_comb_filter}(a). The time $t$ in $f(t)$ is in the range $\SIrange{0}{133.3}{ns}$, i.e., within one IOTA revolution. The number of photoelectrons,
generated by the light pulse at the $i$th turn, $\N_i$, can be calculated as the
time integral of the output pulse of the integrator divided by the electron
charge $e$ and the resistance $\Rf$, i.e.,
\begin{equation}\label{eq:NifromAiInt}
    \N_i=\int A_if(t)\dd{t}/(e\Rf). 
\end{equation}
\noindent The function $f(t)$ is known --- it was measured with a fast
oscilloscope. It was practically the same during all of our measurements,
because $f(t)$ is rather wide (about \SI{30}{ns} FWHM) and the length of the
input light pulses was much smaller (about \SI{2}{ns} FWHM); moreover, the shape of input pulses did
not change significantly. Therefore, during all of our measurements
\begin{equation}\label{eq:AtoN}
    \N_i = \chi A_i,
\end{equation}
\noindent where
\begin{equation}
    \chi = \int f(t)\dd{t}/(e\Rf) = \valAmpToPhotoel,
\end{equation}
\noindent
with a $\pm\SI{5}{\percent}$ uncertainty, because of the uncertainty on $\Rf$.
We verified \Cref{eq:AtoN} empirically at different voltage amplitudes $A_i$ and
different bunch lengths, which define the lengths of the input light pulses. During our experiments at different beam currents, $A_i$ was in the range between $\SI{0}{V}$ and $\SI{1.2}{V}$.

Since we also knew the empirical linear relation between the beam current and
the integrator voltage amplitude, we could use it in \Cref{eq:AtoN} to find the
average number of detected photons (photoelectrons) per one electron of the
electron bunch. The result of this calculation was $\valphotonFluxMeas$. This value is quite close to the result
obtained in our simulation, 
\begin{multline}
    \sum\limits_{s=1,2}\int \dd{k}\dd[2]{\phiv}\etaf{\phiv}\abs{\Ef{\phiv}{}{s}}^2\\
    =\valphotonFluxTheor.
\end{multline}

In our experiment, the expected relative fluctuation of $A_i$ was \SI{e-4}{}--\SI{e-3}{} (rms), which is considerably lower than the
digitization resolution of our 8-bit broad-band oscilloscope. To overcome
this problem, we used a passive comb (notch) filter \cite{smith2010physical}, shown in
Fig.~\ref{fig:integrator_and_comb_filter}(b). The signal splitter divides the
integrator output into two identical signals. The lengths and the
characteristics of the cables in the two arms were chosen such that one of the
signals was delayed by exactly one IOTA revolution and, at the same time, the
losses and dispersion in both arms were approximately equal. The time delay in the comb filter could
be adjusted with a $\SI{0.1}{ns}$ precision. Therefore, the time delay error was
negligible, because the pulses at the entrance of the comb filter were about $\SI{30}{ns}$ long (FWHM).
Finally, a passive hybrid \cite{hybrid} generated the difference and the sum
of the signals in the two arms --- its output channels \D~and \S, respectively.
For an ideal comb filter,
\begin{align}
    \Dit{} =  \xi(A_i-A_{i\shortminus 1})\ft{},\label{eq:Deq}\\
    \Sit{} = \xi(A_i+A_{i\shortminus 1})\ft{}\label{eq:Seq},
\end{align}
\noindent where we assume that the pulse shape of input and output
signals of the comb filter is the same --- $f(t)$. This means that we assume a negligible
dispersion in the comb filter, which is a very good approximation according to
our comparison of input and output pulses with the oscilloscope. Also, as a
result of this comparison, we determined the parameter $\xi=\valxi$. Of course,
our comb filter was not perfect. There was some cross-talk between \D- and
\S-channels, some noise in the signals, a small undesirable reflection in one of
the arms, resulting in a small satellite pulse about $\valtr$ away from the main
pulse, see Fig.~\ref{fig:integrator_and_comb_filter}(c). In addition, the hybrid was AC-coupled.

With these effects taken into account \Cref{eq:Deq,eq:Seq} take the form

\begin{multline}
    \Dit{} = \xi(A_i-A_{i\shortminus 1})\ft{}+\muD \Sit{}
    \\
    +\delta_rA_if(t-t_r)
    +\nuDit
    -\DAC\label{eq:Dit},
\end{multline}
\begin{multline}
    \Sit{} = \xi(A_i+A_{i\shortminus 1})\ft{}+\muS \Dit{}
    \\
    +\delta_r A_if(t-t_r)
    +\nuSit
    -\SAC\label{eq:Sit},
\end{multline}
\noindent where $t$ is within one IOTA turn (\SIrange{0}{133.3}{ns}), $\muD$ and $\muS$ describe the cross-talk between \D- and
\S-channels ($<\SI{1}{\percent}$), $\delta_r A_if(t-t_r)$ describes the reflected pulse in one of the arms (perhaps the short one), $t_r =
\valtr$, $\delta_r\approx\SI{1.5e-3}{}$;  
and it is assumed that the noise contributions $\nuDit$ and $\nuSit$ enter the
equations as sum terms, independent of the signal amplitude; the constants
$\DAC$ and $\SAC$ come from the fact that the hybrid is AC-coupled and the
averages of $\Dit{}$ and $\Sit{}$ over a long time have to be zero.

For each measurement, we recorded $\valwfLen$-long waveforms (about
$\nrev=\valNrev$ IOTA revolutions) of \D- and \S-channels with the oscilloscope
at $\SI{20}{GSa/s}$. The beam current decay was negligible during this $\valwfLen$ acquisition period.

In \Cref{eq:Sit}, the noise, the cross-talk term, and the reflection term are
negligible. The \S-channel can be used to measure the photoelectron count mean $\av{\N}$
during the $\valwfLen$. Using \Cref{eq:AtoN} and the non-negligible part of
\Cref{eq:Sit}, 
\begin{equation}\label{eq:avNfromSigma}
    \av{\N} = \chi\frac{\av{\Sigma(\tpeak)}+\SAC}{2\xi},
\end{equation}
\noindent where we introduced $\tpeak$ --- the time within each turn,
corresponding to the peak of the signal, $f(\tpeak)=1$,
\begin{equation}
    \av{\Sigma(\tpeak)} = \frac{1}{\nrev}\sum\limits_{i=1}^{\nrev}\Sigma_i(\tpeak).
\end{equation}

The idea of using a comb filter is that, in the ideal case, see
\Cref{eq:Deq}, the \D-channel would provide the exact difference between two
consecutive turns in IOTA. In this case we would be able to look directly at the
turn-to-turn fluctuations. The offset would be removed, and the oscilloscope
could be used with the appropriate scale setting, with negligible digitization
noise. In our non-ideal comb filter, see \Cref{eq:Dit}, the additional terms
have some impact on the \D-signal, see Fig.~\ref{fig:integrator_and_comb_filter}(c).
Nonetheless,
by analyzing the \D-signal in a special way, described below, it is possible to determine $\var{\N}$ with sufficient precision.

\begin{figure}[!h]
    \includegraphics[width=1.0\columnwidth]{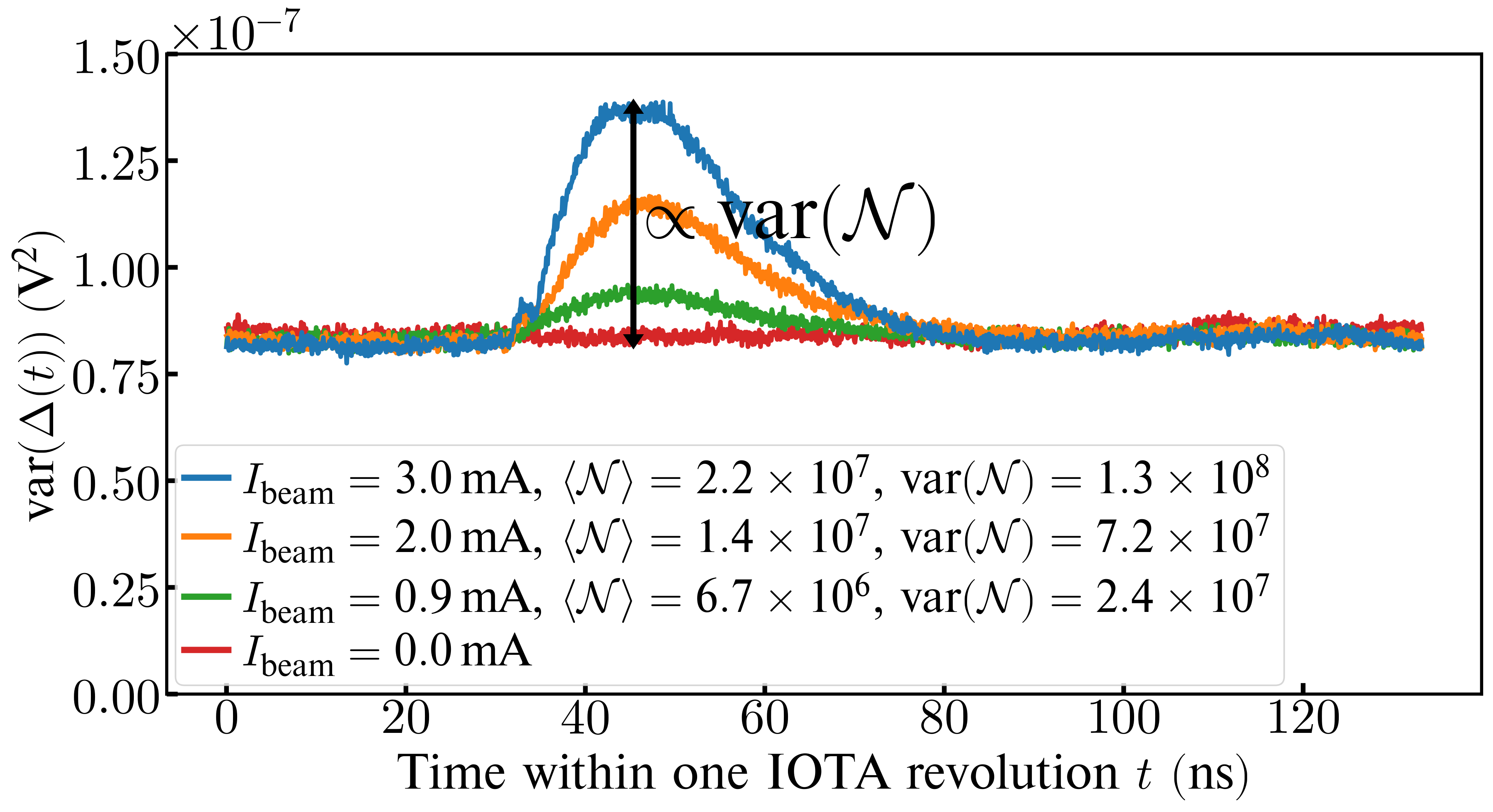}
    \caption{\label{fig:noise_filtering} The variance of \D-signal as a function of
    time (see \Cref{eq:varDt}) within one IOTA revolution (round-beam data).}
\end{figure}

Namely, if we take the variance of \Cref{eq:Dit} with respect to $i$ at a fixed time $t$, then we obtain
\begin{equation}\label{eq:varDt}
    \var{\Dt} = 2\xi^2\var{A}\ft{2}+\var{\nuD(t)},
\end{equation}
\noindent where the contribution from $\muD \Sit{}$ and $\delta_rA_if(t)$ may
be dropped, because the fluctuations of $\Sit{}$ and $A_i$ are strongly
attenuated by the factors $\muD$ and $\delta_r$, respectively. Also,
$\var{\DAC}=0$ since $\DAC$ is constant during the $\valwfLen$. The left-hand side of
\Cref{eq:varDt}, as a function of $t\in[0,\valrevolutionPeriodDimLess]\SI{}{ns}$, could be
obtained from the collected waveforms of \D-channel as
\begin{equation}\label{eq:varDtfromWF}
    \var{\Dt}
    = \frac{1}{\nrev}\sum\limits_{i=1}^{\nrev}\Dit{2}
    - \bigl[\frac{1}{\nrev}\sum\limits_{i=1}^{\nrev}\Dit{}\bigr]^2.
\end{equation}
The results of such calculation for 2000 moments of time $t$ within an IOTA
revolution are shown in Fig.~\ref{fig:noise_filtering}. These data are for the round beam. The blue, orange, and green lines
correspond to three significantly different values of beam current within the
range studied in our experiment; the red line corresponds to a zero beam current case.

Figure~\ref{fig:noise_filtering} suggests that there is a constant noise level,
independent of time and independent of the signal amplitude. Specifically, it
suggests that the noise term in \Cref{eq:varDt} is
\begin{equation}
    \var{\nuD(t)} = \var{\nuD} = \SI{8.8e-8}{V}^2.    
\end{equation}
The observed rms noise amplitude ($\approx\SI{0.3}{mV}$) was analyzed by using the
noise model for the detector electrical schematic,
Figs.~\ref{fig:integrator_and_comb_filter}(a) and (b), as well as the typical
electrical characteristics of the photodiode~\cite{hamamatsu_photodiode} and the
operational amplifier~\cite{opamp}. The three main contributions to the rms
noise in the \D-channel are the following: the oscilloscope input amplifier
noise, \SI{0.21}{mV}; the operational amplifier input voltage noise,
\SI{0.18}{mV}; and the operational amplifier input current noise,
\SI{0.037}{mV}. When added in quadrature, these three sources explain the measured noise level.

The peaks rising above the noise level in Fig.~\ref{fig:noise_filtering} can be
fitted well with $f^2(t)$ (fits not shown). Thus, their shape is in agreement
with \Cref{eq:varDt} as well.

Therefore, using \Cref{eq:AtoN,eq:varDt}, the photoelectron count variance
$\var{\N}$ can be determined as
\begin{equation}\label{eq:varNfromDelta}
    \var{\N} = \chi^2\var{A}
    = \chi^2\frac{\var{\Delta(\tpeak)}-\var{\nuD}}{2\xi^2},
\end{equation}
\noindent see \Cref{eq:varDtfromWF} for the definition of $\var{\Delta(\tpeak)}$.
The value of the noise level term in \Cref{eq:varNfromDelta} is
\begin{equation}
    \frac{\chi^2 \var{\nuD}}{2\xi^2} = \valnoiseLevel.
\end{equation}

We employed a dedicated test light source with known fluctuations to verify this method of measurement of $\av{\N}$ and $\var{\N}$ [\Cref{eq:avNfromSigma,eq:varNfromDelta}]. This verification is described in 
Appendix~\ref{app:testlightsource}, where we also estimated the statistical
error of the measurement of $\var{\N}$ by our apparatus, namely,
$\pm\valvarNerror$ --- it is approximately constant in the range of $\var{\N}$
observed with the undulator radiation in IOTA.

\section{Measurement results}

\begin{figure*}[t]
    \includegraphics[width=\textwidth]{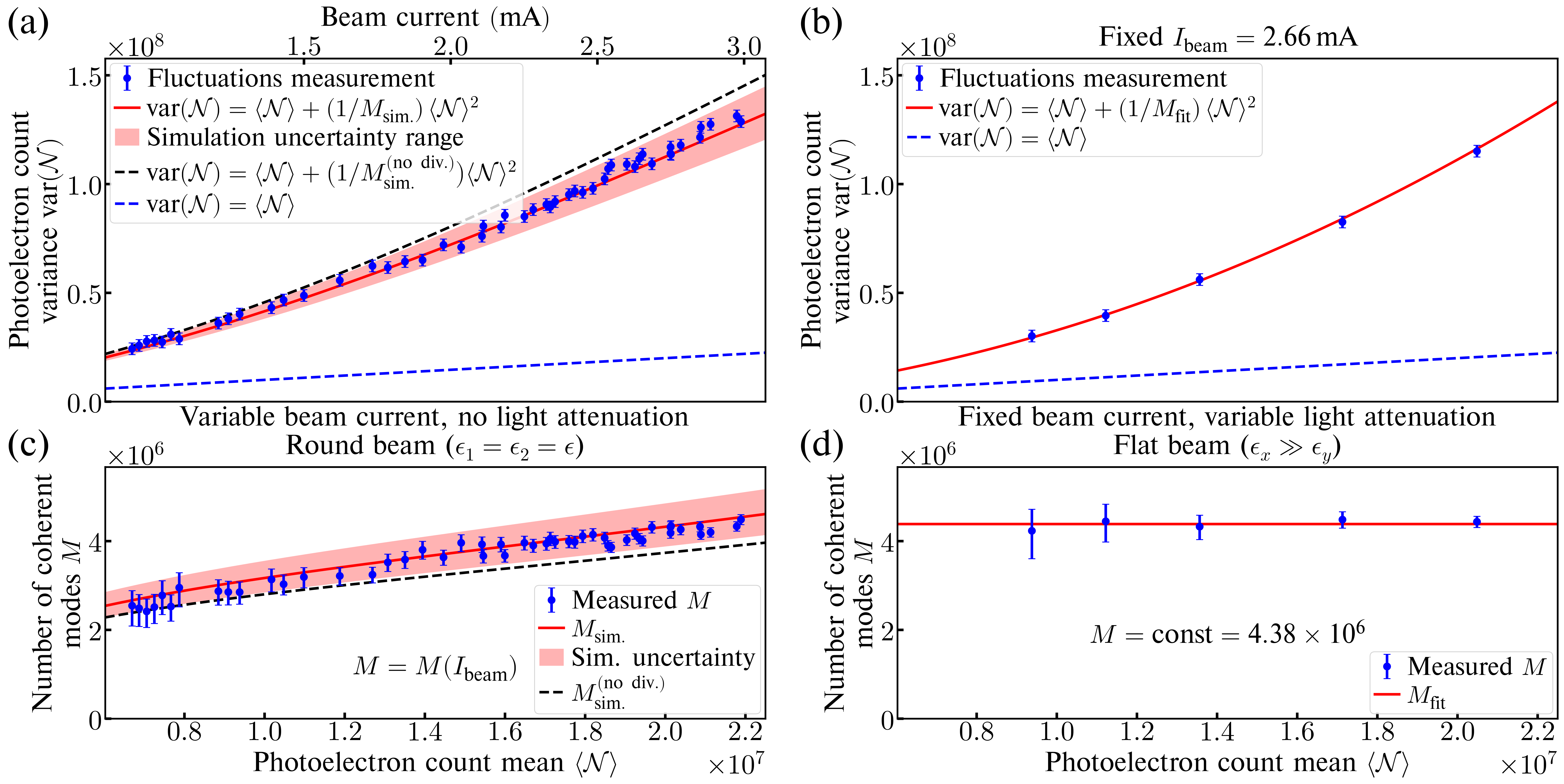}
        \caption{\label{fig:round_beam_compare_with_no_beam_div} 
        (a) The fluctuations measurement for the round electron beam in IOTA as a
        function of beam current, a prediction by \Cref{eq:Mgeneral} (red solid
        line) and a prediction by \MnoDiv~(black dashed line), which does not
        account for the beam divergence. (b) The fluctuations measurement for the flat
        electron beam at a fixed beam current $\valIbeamND$ with 4 different
        optical neutral density filters and one point without any filters, as well as a parabolic fit (a prediction
        could not be made, because the vertical emittance of the flat beam was unknown). (c) and (d) present the data of (a) and (b) in terms of the number of coherent modes $M$.}
\end{figure*}

The measured fluctuations data for the round beam at different values of beam current are shown in Fig.~\ref{fig:round_beam_compare_with_no_beam_div}(a)
(blue points).
The blue dashed straight line, $\var{\N}=\av{\N}$, represents the
photon shot noise contribution to the fluctuations --- the first sum term in
\Cref{eq:varN_from_book}.  The values of $M$ extracted from the fluctuation data
points using the equation,
\begin{equation}\label{eq:MfromvarN}
    M = \av{\N}^2/(\var{\N}-\av{\N}),
\end{equation}
\noindent are shown in Fig.~\ref{fig:round_beam_compare_with_no_beam_div}(c)
(blue points). The error bars in Figs.~\ref{fig:round_beam_compare_with_no_beam_div}(a),(c) correspond to the $\pm\valvarNerror$ statistical error of measurement of $\var{\N}$ by our technique.
Further, Fig.~\ref{fig:round_beam_compare_with_no_beam_div}(c) has a
curve for $M$, simulated by \Cref{eq:Mgeneral} (red line), and, for comparison,
a curve for $M$, simulated by \MnoDiv~(dashed black line), which neglects beam
divergence. Corresponding curves for simulated $\var{\N}$ are shown in
Fig.~\ref{fig:round_beam_compare_with_no_beam_div}(a). The shaded light red
areas in Figs.~\ref{fig:round_beam_compare_with_no_beam_div}(a),(c) show the
uncertainty range of our simulation by \Cref{eq:Mgeneral}.

For this simulation, we needed the values of the following four bunch
parameters, entering \Cref{eq:Mgeneral}, $\ex$, $\ey$, $\sp$, $\szeff$, at all
beam currents. Further, we needed the values of Twiss functions in the
undulator, the parameters of the undulator and of the detection system. All
these aspects were described in Sec.~\ref{sec:apparatus}. We had no free
parameters in this simulation. Numerical calculation of the integrals in
\Cref{eq:Mgeneral} and \MnoDiv~was performed by the Monte-Carlo algorithm on the
Midway2 cluster at the University of Chicago Research Computing Center using our
computer code \cite{furrepo,wigradrepo}.

The simulation uncertainty range (shaded light red area) primarily comes from
the uncertainty in the beam energy $\valEbeamDimLess\pm\valEbeamerror$. The next
source of uncertainty by magnitude, which is a factor of two smaller, is the SLMs' $\pm\valSLMroundErr$
monitor-to-monitor variation of $\e$ [for the round beam we use $\ex=\e$ and $\ey=\e$ in \Cref{eq:Mgeneral}]. The uncertainties of other parameters ($\sp$, $\szeff$, Twiss functions
in the undulator, etc.) had negligible effect. The manufacturers' specifications
for the optical elements of our system did not provide any uncertainties.
Therefore, they were not considered.

We also collected fluctuations data for another experiment configuration.
Figure~\ref{fig:round_beam_compare_with_no_beam_div}(b) shows fluctuation data
points for a flat beam (uncoupled focusing) at a fixed beam current $\valIbeamND$.
The corresponding reconstructed values of $M$ are shown in
Fig.~\ref{fig:round_beam_compare_with_no_beam_div}(d).
The error bars in Figs.~\ref{fig:round_beam_compare_with_no_beam_div}(b),(d) correspond to the $\pm\valvarNerror$ statistical error of measured $\var{\N}$.
In this measurement, the
photoelectron count mean (horizontal axis) was varied by using different optical
neutral density filters (one point without a filter and four points with
filters). Neutral density filters are filters that have constant attenuation in
a certain wavelength range; in our case, around the fundamental harmonic of the
undulator radiation. A new bunch was injected into the ring for each
measurement. The oscilloscope waveforms for \D- and \S-channels were recorded
when the beam current decayed to $\valIbeamND$. The red curve in
Fig.~\ref{fig:round_beam_compare_with_no_beam_div}(b) is a fit with a constant
$M$. A corresponding horizontal line is shown in
Fig.~\ref{fig:round_beam_compare_with_no_beam_div}(d). The value of $M$ in this
fit is $\Mfit=(4.38\pm 0.10)\times 10^6$. This value was calculated as the
average of the five values of $M$ in
Fig.~\ref{fig:round_beam_compare_with_no_beam_div}(d), and the error was
calculated as the standard deviation of these five values.

We do not present any simulation results for the fluctuations $\var{\N}$ in the uncoupled focusing, because
the SLMs provided very inconsistent estimates for the small vertical
emittance $\ey$ of the flat beam --- the max-to-min variation for different SLMs reached a factor of eight.
We believe this happened because the beam images were close to the resolution limit, set by a combination of factors, such as the diffraction limit, the point spread function of the cameras, chromatic aberrations, the effective radiator size of the dipole magnet radiation ($\approx\SI{20}{\micro m}$), and the camera pixel size ($\approx\SI{10}{\micro m}$ in terms of beam size).
Therefore, the monitor-to-monitor emittance value variation primarily came from the Twiss beta-function variation ($\beta_y^{(\mathrm{max})}/\beta_y^{(\mathrm{min})} \approx 12$).
The diffraction limit is primarily caused by the irises, used to reduce the radiation intensity to prevent the cameras from saturating at high beam currents. Alternatively, leaving the irises open and using attenuating optical filters may improve the resolution.
Additional negative effects include the errors in the light focusing optics, calibration errors of the SLMs, and possible Twiss beta-function errors.
The SLMs at locations with larger beta-functions (M4L, M1L) provide estimates for $\e_y$ that agree better with the theoretical predictions \cite{lebedev2020ibs} at lower beam currents, and with the emittance estimates presented below in this paper. Nevertheless, we cannot state that the SLMs could provide a reliable estimate for $\e_y$ during our experiment.

Without $\ey$ we could not use \Cref{eq:Mgeneral} to make a
prediction for $M$ and $\var{\N}$.
However, we attempted the reverse of this procedure, namely, the reconstruction of the
unknown $\ey$ via the measured fluctuations for the flat beam shown in
Figs.~\ref{fig:round_beam_compare_with_no_beam_div}(b),(d). Indeed, the measured
value of the number of coherent modes $\Mmeas$ is a function of four bunch
parameters,
\begin{equation}\label{eq:eyViaMmeas}
    \Mmeas = M(\ex,\ey,\sp,\szeff).
\end{equation}
\noindent The full form of the right-hand side is given by \Cref{eq:Mgeneral}.
The horizontal emittance $\ex$ of the flat beam at a beam current of
$\valIbeamND$ could still be reliably measured via the SLMs, yielding $\e_x=\valexND$. The
effective bunch length $\szeff$ could be determined from $\rho(z)$
measured by the wall-current monitor, $\szeff=\valszeffND$. The rms momentum
spread $\sp$ was estimated from $\rho(z)$ and the ring parameters,
$\sp=\valdppND$. The only unknown in \Cref{eq:eyViaMmeas} is $\ey$.
Equation~\eqref{eq:eyViaMmeas} can be solved for $\ey$ by a simple bisection
method. The result is $\ey = \valeyWithDivDimLess \pm \valeyWithDivErr$, where
the uncertainty corresponds to the statistical error of $\Mfit$, mentioned
above. For comparison, we also used \MnoDiv~ in \Cref{eq:eyViaMmeas}, which
neglects beam divergence. In this case, we obtained $\ey =
\valeyWithoutDivDimLess \pm \valeyWithoutDivErr$.

In this reconstruction of $\ey$, there is also a
systematic error due to the uncertainty on the beam energy ($\pm\valEbeamerror$)
and due to the systematic error of $\ex$ measurement by the SLMs ($\pm\valSLMflatErr$
monitor-to-monitor). We estimated these two contributions to the systematic
error of $\ey$ from \Cref{eq:Mgeneral}. They are
$\substack{+2.5 \\ -4.5}\;\SI{}{nm}$ and $\substack{+1.6 \\ -1.4}\;\SI{}{nm}$,
respectively. These systematic errors are rather significant. However, they are
not directly linked to our measurement technique. They are related to the fact
that the beam energy and the horizontal emittance of a flat beam in IOTA were not
known with better precision. Further improvements in beam
characterization in IOTA will reduce the systematic error of our
fluctuations-based technique of $\ey$ measurement.

\section{\label{sec:discussion}Discussion and conclusions}

Power fluctuations in undulator radiation were measured in IOTA under two different experimental conditions and compared with our theoretical predictions.

The fluctuations predicted by \Cref{eq:varN_from_book,eq:Mgeneral} in the round beam configuration agree with the measurements within the uncertainties, as shown in Figs.~\ref{fig:round_beam_compare_with_no_beam_div}(a),(c).
In IOTA, the bunch parameters $\ex$, $\ey$, $\sp$,
$\szeff$ depend on the beam current because of various intensity dependent
effects, e.g., intra-beam scattering \cite{reiser1994theory}, beam interaction
with its environment \cite{haissinski1973exact}, etc. Therefore, $M$ is a
function of the beam current as well, as one can see in
Fig.~\ref{fig:round_beam_compare_with_no_beam_div}(c).

In Figs.~\ref{fig:round_beam_compare_with_no_beam_div}(b),(d), all data points
correspond to one value of the beam current, $\valIbeamND$. The photoelectron count
mean is varied by using neutral density filters with different attenuation
factors \newcommand{\etaND}{\eta_{\mathrm{ND}}} $\etaND$. Such filters linearly
scale down the photoelectron count mean, $\av{\N}\rightarrow\etaND\av{\N}$.
However, they do not change $M$, because if $\etaf{\phiv}$ is replaced by
$\etaND\etaf{\phiv}$ in \Cref{eq:Mgeneral}, then $\etaND$ cancels out in the
numerator and the denominator. This is consistent with
Fig.~\ref{fig:round_beam_compare_with_no_beam_div}(d) --- all measured
values of $M$ are equal within the uncertainty range.

We reconstructed the value
of the vertical emittance $\ey$ of the flat beam from $\Mfit$ via \Cref{eq:Mgeneral}
and via \MnoDiv, which neglects beam divergence. We obtained very different
results, $\ey = \valeyWithDivDimLess \pm \valeyWithDivErr$ and $\ey =
\valeyWithoutDivDimLess \pm \valeyWithoutDivErr$, respectively. This shows that,
for the flat beam, accounting for the beam divergence is critical. In this
measurement, the horizontal beam divergence was $\valbeamDivXND$ and comparable with the rms radiation divergence $\sqrt{\lambda_1/(2\Lu)}=\valoneOverGammaNu$ \cite[Eq.~(2.57)]{kim2017synchrotron},
which gives an estimate of the angular size of $\Ef{\phiv}{}{s}$. Clearly, in
this case it has a significant effect on the integral in the numerator of
\Cref{eq:Mgeneral}. This is why we use \Cref{eq:Mgeneral} in our Letter
\cite{lobach2020furjointprl}, focused on emittance measurements via fluctuations,
as opposed to \MnoDiv, which is simpler, but neglects beam divergence.

In addition, we made an independent estimate of the vertical emittance $\ey$ of
the flat beam based on the beam lifetime, see Appendix~\ref{app:touschek} for
details. At the beam current of $\valIbeamND$, the beam lifetime is solely
determined by Touschek scattering \cite{lebedev2020ibs,lebedev2010betatron,touschek}. The Touschek lifetime is a function of beam
emittances and bunch length. Therefore, since we knew the horizontal emittance $\ex$ of the flat
beam, measured by the SLMs, the bunch length, measured by the wall-current monitor, and the measured beam lifetime
$\abs{I/(\dv*{I}{t})}$, we could find $\ey$. The result is $\ey=\valeyLifetimeRecNDDimLess\pm\valeyLifetimeRecNDErr$, to be compared with the fluctuations-based
measurement, $\ey = \valeyWithDivDimLess \pm \valeyWithDivErr$.
The $\pm\valeyLifetimeRecNDErr$ error in the lifetime-based $\ey$ estimate comes from the $\pm\valSLMflatErr$ uncertainty on $\ex$ of the flat beam.

In the round beam case, at the same beam current of $\valIbeamND$, the beam divergence in the
undulator was about $\valbeamDivRoundX$ (both $x$ and $y$). It was noticeably smaller than the
horizontal beam divergence of the flat beam and than
the rms radiation divergence $\valoneOverGammaNu$. Therefore, the effect of
beam divergence on the fluctuations simulation in
Figs.~\ref{fig:round_beam_compare_with_no_beam_div}(a),(c) is not as dramatic.
However, the deviation from the measurement of the simulation based on \MnoDiv, which neglects beam divergence, is certainly noticeable, whereas the simulation by
\Cref{eq:Mgeneral} agrees well with the measurement.

It would be beneficial to repeat these fluctuation measurements with a
longer and brighter undulator.
In this experiment, we had to avoid using a
monochromator or restricting the angular aperture, because we had to collect all available radiation to
achieve a signal with a voltage amplitude that could be easily measured. Therefore, the integrals in \Cref{eq:Mgeneral,eq:NoneDef} had to be
calculated over a broad range of angles and wavelengths. With a
monochromator, a slit or a pinhole detector, these integrals could be significantly simplified. Further, if our undulator had more periods and if  we were able to use a monochromator, we could slowly vary the beam energy and find the energy at which the detected power is at maximum, i.e., when we are
centered on the peak of the fundamental harmonic. In this case, the systematic
error of the $\ey$ measurement via the fluctuations, related to the uncertainty on the beam energy, would
be negligible. Moreover, it can be shown \cite{lobach2020furjointprl} that if one places a narrow vertical slit in front of the detector, then the magnitude of the fluctuations would only depend on $\ey$, i.e., the systematic error of
the $\ey$ measurement related to the
uncertainty of $\ex$, measured by the SLMs, would be minimized, too. Finally, for a
brighter undulator, the statistical error on the measured value of the fluctuations
would be lower as well.

We are considering using our fluctuations measurement apparatus, or an improved variation of it, as
a tool for the diagnostics of the Optical Stochastic Cooling (OSC) experiment in IOTA
\cite{lebedev2015optical,jarvis2018optical,andorf2018computation}. In this experiment, the beam emittance will be even smaller and the existing diagnostic tools may not have sufficient resolution. Even though we cannot measure
transverse emittances and bunch lengths individually in this way, the
fluctuations may serve as an indicator of the cooling process. When the
cooling process starts, the electron bunch shrinks, which, in turn, makes the
fluctuations of the number of detected photons increase.

To conclude, we presented a calculation [\Cref{eq:varN_from_book,eq:Mgeneral}], which describes the turn-to-turn fluctuations in the number
of detected synchrotron radiation photons, produced by an electron bunch with a Gaussian transverse density distribution, an arbitrary longitudinal
density distribution, and non-negligible rms beam divergences.
The rms bunch
length is assumed to be significantly larger than the radiation wavelength. Equation~\eqref{eq:Mgeneral} is
presented for the first time, as beam divergence has been neglected in all previous considerations \cite{sannibale2009absolute,lobach2020PRAB}. Beam divergence can
be neglected if it is significantly smaller than the characteristic radiation
angle --- $\sqrt{\lambda_1/(2\Lu)}$ for the fundamental of undulator radiation. We
presented the results of an experiment with a round beam in IOTA, where beam
divergence had an impact on the fluctuations of the undulator radiation. We
showed that our new \Cref{eq:Mgeneral} agrees better with the measurements than
\MnoDiv, which neglects beam divergence. Finally, we proposed a non-invasive
technique to measure the small unknown vertical emittance of a flat beam via the fluctuations. This new technique is described in detail in a separate publication \cite{lobach2020furjointprl}.

\begin{acknowledgments}
We would like to thank the entire FAST/IOTA team at Fermilab for helping us with
building and installing the experimental setup and taking data, especially Mark
Obrycki, James Santucci, and Wayne Johnson. Greg Saewert constructed the detection circuit and provided the test light source.
Brian Fellenz, Daniil Frolov, David Johnson, and Todd Johnson provided equipment and assisted during our
detector tests.
This work was completed in part with resources provided by the University of Chicago Research Computing Center. This research is supported by the University of Chicago and the US Department of Energy under contracts DE-AC02-76SF00515 and
DE-AC02-06CH11357. This manuscript has been authored by Fermi Research Alliance,
LLC under Contract No. DE-AC02-07CH11359 with the U.S. Department of Energy,
Office of Science, Office of High Energy Physics.
\end{acknowledgments}

\appendix

\section{\label{app:beamdiv}Derivation of the fluctuations with a considerable beam divergence}

Previously, we derived an equation for $M$ \MnoDiv~for the case of a
monoenergetic beam, zero beam divergence and temporally incoherent radiation.
Below we outline the steps to extend this result to the case of a considerable
beam divergence [see \Cref{eq:Mgeneral}].

One can start from \cite[Eq.~(21)]{lobach2020PRAB}, but written in a form that
accounts for the beam divergence, namely,
\begin{multline}\label{eq:varN1}
    \var{\N} = \av{\N}-\av{\N}^2
    \\
    +\intens\Bigl[
        \sum\limits_{s=1,2}
        \int\dd{k} \dd[2]{\phiv} \etaf{\phiv} \bigl|\sum\limits_m \Emf{\phiv}\bigr|^2
    \Bigr]^2,
\end{multline}
\noindent where $\ens$ describes the states in the 6D phase-space of all
the electrons in the center of the radiator, 
\begin{multline}\label{eq:enssixdim}
    \ens = \sixcoor{1}
    \\
    \dots
    \sixcoor{\Ne},
\end{multline}
\noindent where $t_m$ is the time when the $m$th electron passes the center of the synchrotron light source, $\pens$ represents the density function for the probability to have the state
$\ens$,
\begin{multline}\label{eq:penssixdim}
    \pens = \rhobunch{\sixcoor{1}}
    \\
    \dots
    \rhobunch{\sixcoor{\Ne}},
\end{multline}
\noindent $\monx_m$ and $\monxp_m$ refer to the monoenergetic component of the
motion, because there is also a contribution from the horizontal dispersion, so
that
\begin{align}\label{eq:fullxmxpm}
    & x_m = \monx_m+\Dx \dpp_m, && \xp_m = \monxp_m+\Dxp \dpp_m,
 \end{align}
\noindent and the vertical dispersion is assumed to be zero. According to \cite[Eq.~(2.93)]{kim2017synchrotron}, the complex field
amplitude of the $m$th electron $\Emf{\phiv}$ can be expressed through the amplitude
of the reference electron $\Ef{\phiv}{}{s}$ [see \Cref{eq:Adef}] as
\begin{equation}\label{eq:EmViaE}
    \Emf{\phiv} = e^{ict_m-ik_xx_m-ik_yy_m}\Ef{\phiv-\xpv_m}{}{s},
\end{equation}
\noindent where $\xpv_m = (\xp_m, \yp_m)$ and it is assumed that $\Emf{\phiv}$
does not depend on $\dpp_m$. For the reference electron $\sixcoor{}$ are equal to zero.

We assume an electron bunch that is Gaussian in the transverse plane and has a
Gaussian distribution in $\dpp$. The longitudinal density distribution
$\rho(z)$ is arbitrary. The beam focusing optics is assumed to be uncoupled. In this case, the probability density function for one electron
takes the following form,
\begin{widetext}
    \begin{equation}\label{eq:psixdim}
        \rhobunch{\sixcoor{}}
            = \frac{1}{4\pi^2\ex\ey}
            \exp\Bigl[
                -\frac{1}{2\ex^2}C_x(\monx,\monxp)-\frac{1}{2\ey^2}C_y(y,\yp)
                \Bigr]
            \rho(-ct)
            \frac{1}{\sqrt{2\pi}\sp}
            \exp\Bigl[
                -\frac{\dpp^2}{2\sp^2}
                \Bigr],
    \end{equation} 
\end{widetext}
\noindent with
\begin{align}
    C_x(\monx,\monxp) = \gx \monx^2 + 2 \ax \monxp + \bx (\monxp)^2\label{eq:Cx},
    \\
    C_y(y,\yp) = \gy y^2 + 2 \ay \yp + \by (\yp)^2\label{eq:Cy}.
\end{align}

Given
\Cref{eq:enssixdim,eq:penssixdim,eq:fullxmxpm,eq:EmViaE,eq:psixdim,eq:Cx,eq:Cy},
and assuming the regime of longitudinal incoherence ($k\sz\gg1$), the
integration in \Cref{eq:varN1} is solely a mathematical procedure. It is
analogous to the derivation in \cite{lobach2020PRAB} where $\ens$ included only
$x_m$, $y_m$ and $t_m$. The only difference is the additional integration over
$\monxp_m$, $\yp_m$ and $\dpp_m$, with $m=1\dots\Ne$. When the multidimensional
integral in \Cref{eq:varN1} has been calculated, one can compare the result with
\Cref{eq:varN_from_book} and arrive at an expression for $M$ as in
\Cref{eq:Mgeneral}.

\begin{figure*}[t]
    \includegraphics[width=\textwidth]{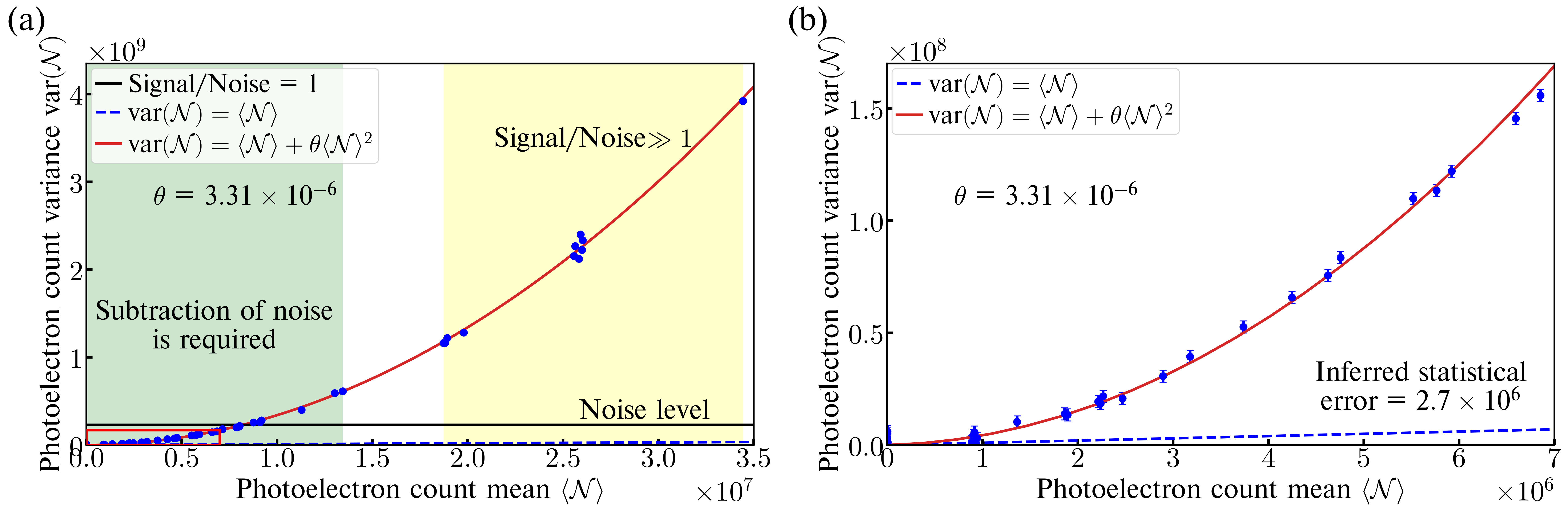}
    \caption{\label{fig:test_light_source} Photoelectron count variance
    $\var{\N}$ as a function of photoelectron count mean $\av{\N}$ for the test
    light source; $\av{\N}$ was varied by using different neutral density
    filters. (a) The entire range of $\av{\N}$, $\var{\N}$. (b) The region
    corresponding to the values of $\var{\N}$ generated by the undulator radiation
    in IOTA [highlighted by the red rectangle in (a)].}
\end{figure*}

\section{\label{app:gaussianlightbeam}Approximation of a Gaussian radiation
profile}

In this approximation, the following expression for the radiation field
amplitude is used,
\begin{widetext}
    \begin{equation}\label{eq:gaussianLightBeam}
        \Ef{\phiv}{}{s} = \sqrt{\frac{C_s}{(2\pi)^{3/2}\usk\usxp\usyp}\exp\bigl
        [-\frac{(k-k_0)^2}{2\usk^2}-\frac{\f{x}^2}{2\usxp^2}-\frac{\f{y}^2}
        {2\usyp^2}\bigr]},
    \end{equation}
\end{widetext}
\noindent where $k_0$ refers to the center of the spectrum of the Gaussian
radiation, $\usk$ is the spectral rms width, $\usxp$ and $\usyp$ are the angular rms
radiation sizes, $\usk\ll 1/(\sx\usxp)$ and $\usk\ll 1/(\sy\usyp)$, $C_s$ is a constant. An ideal
detector is assumed --- $\etaf{\phiv}=1$. In this case, for a Gaussian electron
bunch, the following result can be obtained from \Cref{eq:Mgeneral}, 
\begin{multline}\label{eq:Mgauss}
    M = (1-1/\Ne)^{-1}
    \sqrt{1+4\usk^2\sz^2}
    \\
    \times\sqrt{1+4\kzero^2 (\sx^2 \usxp^2
        + \sxp^2 \Sx^2)
            + \frac{\sxp^2}{\usxp^2}}
    \\
    \times\sqrt{1+4\kzero^2 (\sy^2 \usyp^2
        + \syp^2 \Sy^2)
            + \frac{\syp^2}{\usyp^2}}.
\end{multline}

In the limit of zero electron beam divergence ($\sxp,\syp=0$), \Cref{eq:Mgauss}
coincides with \cite[Eq.~(17)]{sannibale2009absolute}, where this less general
case was considered.

\section{\label{app:testlightsource}Measurements with a test light source}

The method of determining $\av{\N}$ and $\var{\N}$ by
\Cref{eq:avNfromSigma,eq:varNfromDelta} was tested with an independent test
light source with known fluctuations. The test light source consisted of a fast
laser diode ($\SI{1064}{nm}$) with an amplifier, modulated by a pulse generator.
The width of the light pulses and the repetition rate were very close to the
experiment conditions in IOTA. However, the pulse-to-pulse fluctuations in the
test light source were significantly greater than in the undulator radiation in
IOTA, namely, $\var{\N}=\SI{4e9}{}$ as opposed to
$\var{\N}=\SIrange{0}{1.5e8}{}$ in IOTA. This also means that they were much
greater than the instrumental noise level of our apparatus, $\valnoiseLevel$. Therefore, we could reliably
measure the relative fluctuations in the test light source, even without subtraction of the noise level, because it was negligible. The result was
\begin{equation}\label{eq:thetaDef}
    \theta = \frac{\var{\N}}{\av{\N}^2} = \valtheta,
\end{equation} 
\noindent which corresponds to the rms value $\valthetarms$. We believe that these
fluctuations primarily came from the jitter in the pulse generator amplitude.

Further, we used neutral density filters to lower the number of photons detected
by our apparatus. Neutral density filters are filters that have constant optical
density in the wavelength region of interest. As they lower $\av{\N}$ for the
test light source, $\var{\N}$ is lowered in the following known way,
\begin{equation}\label{eq:testLightSourceNDfilters}
    \var{\N} = \av{\N} + \theta\av{\N}^2,
\end{equation}
\noindent i.e., the relative fluctuations stay practically constant
$\var{\N}/\av{\N}^2\approx\theta$, because they are caused by the pulse
generator amplitude jitter, but at a very low $\av{\N}$ the photon shot noise
term [the first term in \Cref{eq:testLightSourceNDfilters}] may have a
noticeable contribution, similar to \Cref{eq:varN_from_book}. By using many
different neutral density filters and their combinations we were able to record
\D- and \S-channel waveforms for a wide range of $\var{\N}$, see
Fig.~\ref{fig:test_light_source}(a), including the range observed in our
experiment in IOTA, shown in Fig.~\ref{fig:test_light_source}(b) and highlighted
by a red rectangle in Fig.~\ref{fig:test_light_source}(a).

In Figs.~\ref{fig:test_light_source}(a) and (b), the parameter $\theta$  of the
red predicted curve was obtained in a configuration without any neutral
density filters, when the detector noise and the photon shot noise were
negligible, see \Cref{eq:thetaDef}. The blue fluctuation data points, obtained
from the \D- and \S-channel waveforms using \Cref{eq:avNfromSigma,eq:varNfromDelta}, agree with the
red predicted curve in the entire range of $\var{\N}$, including the range of
Fig.~\ref{fig:test_light_source}(b) corresponding to the measurements in IOTA.
This means that the method of extracting $\var{\N}$ from the waveforms,
described in Fig.~\ref{fig:noise_filtering} and \Cref{eq:varNfromDelta}, works well, and that the instrumental noise
[$\var{\nuD}=\valnoiseLevel$] is indeed independent of the signal amplitude.

We estimated the statistical error of our measurement of photoelectron count
variance in IOTA as the rms deviation of the fluctuation data points for the test light source from the
predicted curve in Fig.~\ref{fig:test_light_source}(b). The error is
$\pm\valvarNerror$. It is used in the error bars in
Figs.~\ref{fig:round_beam_compare_with_no_beam_div},\ref{fig:test_light_source}(b).

\section{\label{app:touschek}Vertical emittance estimation for the flat beam via the Touschek beam lifetime}

\begin{figure}[!h]
    \includegraphics[width=1.0\columnwidth]{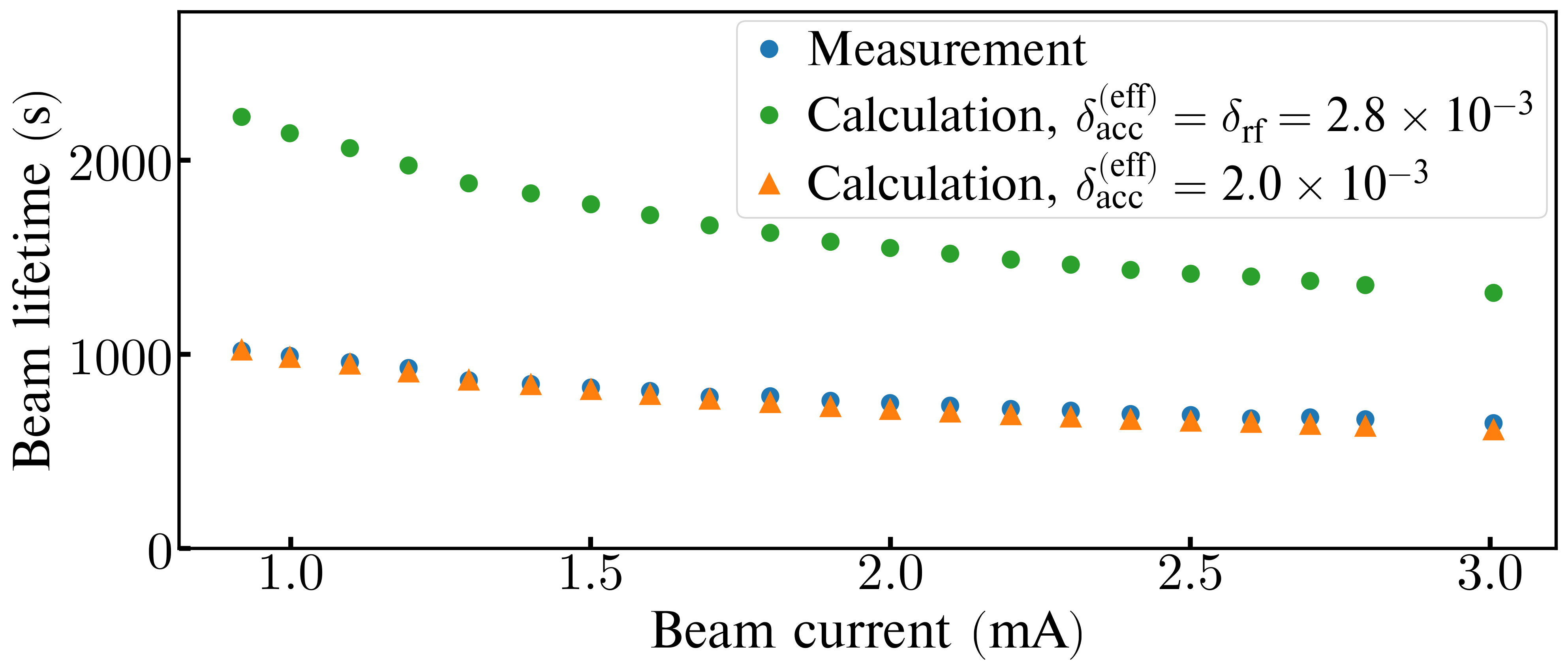}
    \caption{\label{fig:touschek} Lifetime of the round beam in IOTA as a function of beam current.}
\end{figure}

The beam lifetime could be reliably determined from the measured beam current
$I$ as a function of time, as $\abs{I/(\dv*{I}{t})}$. During all of our
measurements the beam current was measured with a DCCT current monitor, it was reported every second. Further, the waveforms from the wall-current monitor allowed us to see the distribution of the electrons among the 4 rf buckets in IOTA. We always corrected the DCCT current to only account for the main bucket. Typically, the combined population of the remaining 3 buckets was no more than a few percent. At the beam currents studied in our experiment
(\SIrange{1}{3}{mA}), the beam lifetime is determined solely by Touschek
scattering \cite{touschek,lebedev2020ibs}. In general, the momentum acceptance is a function of the position
along the ring. It is limited by the longitudinal bucket size, $\momApRF$, and
by the dynamic momentum aperture. A constant effective momentum acceptance
$\momApEff$ can be used \cite{Carmignani:2014qyb} to describe the losses due to Touschek scattering.
It is equal to or smaller than $\momApRF$. 
We used the approach described in
\cite{Intro_touschek,piwinski1999touschek} to calculate the
Touschek lifetime.
Figure~\ref{fig:touschek} shows the measured beam lifetime for the round beam, a calculation with the momentum acceptance limited only by the rf bucket size ($\momApRF=\valdppSep$ in IOTA), and a calculation with an effective momentum acceptance $\momApEff=\valMomApEff$. 

The calculation with $\momApEff=\valMomApEff$ almost perfectly agrees with
the measurement. The emittances and the beam lifetime of the round beam are
known with good accuracy, the only unknown in this Touschek lifetime
calculation for the round beam being $\momApEff$. We believe that Fig.~\ref{fig:touschek} illustrates that in IOTA $\momApEff=\valMomApEff$.

Further, we can also apply this Touschek lifetime calculation (with
$\momApEff=\valMomApEff$) to the flat beam at a beam current of $\valIbeamND$,
where we know the measured lifetime ($\vallifetimeND$) and the horizontal
emittance, but we cannot measure directly the small vertical emittance. We
found the following value for the vertical emittance, $\ey=\valeyLifetimeRecNDDimLess\pm\valeyLifetimeRecNDErr$,
to be compared with the fluctuations-based measurement, $\ey =
\valeyWithDivDimLess \pm \valeyWithDivErr$.
The $\pm\valeyLifetimeRecNDErr$ error in the lifetime-based $\ey$ estimate comes from the $\pm\valSLMflatErr$ uncertainty on $\ex$ of the flat beam.
Other sources of error are much smaller: $\pm\SI{5}{sec}$ uncertainty on the measured beam lifetime, uncertainties in the beam energy, Twiss-functions, and bunch length.

\newpage

\bibliography{bibliography}

\end{document}